\documentclass[epj]{svjour}
\usepackage{graphicx}
\usepackage{dcolumn}
\usepackage{bm}
\usepackage{ctable}
\usepackage{amsmath}
\usepackage{soul,xcolor}
\usepackage{cancel}

\newcommand{\be}{\begin{equation}}
\newcommand{\ee}{\end{equation}}
\newcommand{\beq}{\begin{eqnarray}}
\newcommand{\eeq}{\end{eqnarray}}
\newcommand{\bce}{\begin{center}}
\newcommand{\ece}{\end{center}}

\newcommand{\rmat}{$\mathbf{R}$-matrix }
\newcommand{\etal}{\textit{et al.}}
\newcommand{\comment}[1]{}

\begin{document}
\setstcolor{red}

\title{Higher lying resonances in low-energy electron scattering with carbon monoxide}

\author{Amar Dora\inst{1}, Jonathan Tennyson\inst{2} \and Kalyan Chakrabarti\inst{3}}

%\email{j.tennyson@ucl.ac.uk}
\institute{Department of Chemistry, North Orissa University, Baripada 757003, Odisha, India
\and
Department of Physics and Astronomy, University College London, Gower St., 
London WC1E 6BT, UK \and
Department of Mathematics, Scottish Church College, 1 \& 3 Urquhart Sq, 
Kolkata 700 006, India}

\date{Received: date / Revised version: date}
% The correct dates will be entered by Springer

\abstract{ 
\rmat calculations on electron collisions with CO are reported whose aim is to 
identify any higher-lying resonances above
the well-reported and lowest $^2\Pi$ resonance at about 1.6~eV.
Extensive tests with respect to basis sets, target models and
scattering models are performed.  The final results are reported for
the larger cc-pVTZ basis set using a 50 state close-coupling (CC)
calculation. The Breit-Wigner eigenphase sum and the
time-delay methods are used to detect and fit any resonances.  Both
these methods find a very narrow $^2\Sigma^+$ symmetry Feshbach-type
resonance very close to the target excitation threshold of the b
$^3\Sigma^+$ state which lies at 12.9 eV in the calculations.  This
resonance is seen in the CC calculation using cc-pVTZ
basis set while a CC calculation using the cc-pVDZ basis set does not produce
this feature.  The electronic structure of CO$^-$ is analysed
in the asymptotic region; 45 molecular states are found to 
correlate with states dissociating to an anion and an atom. Electronic
structure calculations are used to study the behaviour of these
states at large internuclear separation. 
Quantitative results for the total, elastic and
  electronic excitation cross sections are also presented. The
  significance of these results for models of the observed dissociative
  electron attachment of CO in the 10 eV region is discussed.  \PACS{
    {34.80.-i}{Electron and positron scattering} %\and
%      {PACS-key}{describing text of that key}
     } % end of PACS codes
} %end of abstract

\authorrunning{Dora, Tennyson and Chakrabarti}
\titlerunning{Higher lying resonances in CO}
\maketitle

\section{Introduction} \label{sec:intro} 

Low-energy electron collisions with the carbon monoxide molecules
display a broad $^2\Pi$ symmetry shape resonance at about 1.6~eV.
This resonance has been well-characterised experimentally
\cite{HM83,Allan89,gmg96,pbv06,allan10} and is the subject of a number of
theoretical studies \cite{gmg96,sbn84,Morgan91,wmz13,jt527}.  This
resonance provides the main mechanism for electron impact vibrational
excitation \cite{jt527,jt621} but lies too low in energy to lead to
dissociative electron attachment (DEA) unless the CO target is also
vibrationally excited \cite{jt621}.

There are a number of theoretical studies which consider collision
energies of above 5 eV; these have generally focused on electron
impact electronic excitation cross sections
\cite{wh90,jt140,lmf96,lib02,gmm05,rtj10}. However, experimentally it
has long been known that CO can undergo dissociative electron
attachment (DEA) with two peaks at about 10.2 and 10.9 eV
\cite{rb65,ss70,cr71,ss71,75cth,hcs77,nn15} and the main product of
this is C+O$^-$, although C$^-$+O has also been observed \cite{ss70}.
The precise physics of the states contributing to the DEA process has
recently proved somewhat controversial \cite{nn15,twx13,nn15b,tl15,wxl15}.

DEA is assumed to occur via resonances. Several works, both experimental
and theoretical,  exist on the resonance features around 
10 eV. In fact, a $^2\Sigma+$ resonance at 10.04 eV  was identified and known 
to contribute to DEA as early as 1973 \cite{Schulz73}. 
 There are  two R-matrix calculations which 
showed resonance features above 20~eV. In one, Salvini {\it et al}
\cite{sbn84} suggested that CO has a $^2\Sigma^+$ shape resonance at
about 20 eV; there is no other evidence for this state. Similarly, in
another calculation, Morgan and Tennyson \cite{jt140} found a rather
broad, widths greater than 1 eV, resonance for each of the doublet
symmetries they considered. Their resonance curves all looked rather
similar. It is at least possible that these features are a consequence
of employing a target wavefunction which extended outside the
R-matrix sphere; this was found to produce artificial resonances in
subsequent calculations on water \cite{mor98,jt281}. Weatherford
and Huo \cite{wh90} performed a two-state calculation using Hartree-Fock
wavefunctions and also found 4 resonances in the 10 - 20 eV region.
 
The available cross section data for
electron collisions with carbon monoxide have recently been collected
and reviewed by Itikawa \cite{Itikawa15}.
These data are important for understanding
discharges, including the CO laser \cite{wc75,GC84,kb15}, other CO
plasmas \cite{14ccd} and a variety of astronomical applications
\cite{lv94,cb09,cab11} as CO is thought to be the second most common
molecule in the Universe after H$_2$.

Given the significance of DEA of CO, it is important to try and build a
viable theoretical model for this process.  As a first step it is
necessary to identify possible resonances through which the DEA may
occur. An aim of this 
paper is to identify  resonances in the 10 -- 12 eV region. In this context 
we note that an 11.5 eV $^2\Sigma_g^+$ resonance has been identified in the
isoelectronic N$_2$ from both experiment and theory \cite{haf09}.
However this theoretical work used bound state methodology to
characterise the resonance, a procedure that is not without dangers
\cite{jt241}. Similarly,  Pearson and Lefebvre-Brion
performed stabilization calculations on the CO$^-$ systems and identified
a single, narrow $^2\Sigma^+$ symmetry resonance at 10.2 eV \cite{pl76}.

\section{Theory} \label{sec:theory}

The general theory of the \rmat method and its specific implementations in the
UK molecular \rmat\ codes, to study various aspects in electron and positron
collision with molecules, has been described in detail in a recent review by
one of us \cite{jt474}. Therefore, we limit ourself to the essential parts that
are necessary for the subsequent discussions.

The \rmat method involves separation of space around the electron+target
collision system depending upon the kind of interaction between the target
molecule and the scattering electron. This separation is usually done using an
imaginary sphere, called the \rmat\ sphere, centred  at the centre-of-mass 
of the molecule. The radius of the sphere is chosen such that it contains
the entire wave function of the $N$-electron target states. Inside the sphere,
called the inner region, the collision complex is described by fully taking
care of the exchange and correlation effects among all the $N+1$ electrons. The
inner region wave function, $\psi_k^{N+1}$, is expressed as a close-coupling (CC) expansion:
\beq 
\begin{split}
\psi_k^{N+1}& = \mathcal{A} \sum_{ij} a_{ijk} \Phi_i^N (\textbf{x}_1...\textbf{x}_N) u_{ij}
(\textbf{x}_{N+1})  \\
&\quad + \sum_i b_{ik} \chi_i^{N+1}(\textbf{x}_1...\textbf{x}_{N+1})\:,\label{eq:rmat} 
\end{split}
\eeq where, in the first term, $\Phi_i^N$ is the wave function of the
$i$-th target state, $u_{ij}$ are the continuum orbitals to represent
the scattering electron and $\mathcal{A}$ is the anti-symmetrization
operator. In the second term, the $\chi_i^{N+1}$ are the so-called
$L^2$ configurations, which are constructed by occupation of all $N+1$
electrons to the target molecular orbitals (MOs).

Different scattering models can be constructed by choosing different
types of expansions for the target wave function ($\Phi_i^N$) and the
corresponding $L^2$ configurations in Eq.~(\ref{eq:rmat}). Generally
three different models are used, namely, the static exchange (SE), SE
plus polarization (SEP) and the close-coupling (CC) models. The SE and
SEP models are among the simplest approximations to the scattering
problem and only use the ground state of the target, represented by a
Hartree-Fock (HF) self consistent field (SCF) wave function. Using the
SE, one can describe only the shape resonances and compute the elastic
cross section. While SEP can represent Feshbach resonances, these are
often not well represented without inclusion of their parent
electronic state.  CC models are more sophisticated and involve
inclusion of several target states which themselves can be represented
by different methods. Usually, the complete active space (CAS)
configuration interaction (CI) method is chosen for representation of
the target states \cite{jt189}. The CC model can describe Feshbach
resonances and also compute electron impact electronic excitation
cross sections. Even more sophisticated is the molecular \rmat\ with
pseudo-states (RMPS) method \cite{jt341,jt354}.  However this method
rapidly leads to huge calculations \cite{jt444} and, given the number
of excited electronic states of the target that need to be explicitly
considered here (see below), a full RMPS study was deemed to be
impractical. A recent attempt to treat several target electronic
states of the simpler, 10-electron methane system shows how large such
calculations rapidly become \cite{jt585}.

At the boundary of the \rmat\ sphere, the \rmat\ is built, for different
scattering energies, from the boundary amplitude of the inner region
wave functions and the \rmat\ poles. Then, the \rmat\ is propagated to large
distances in order to match to the analytical asymptotic functions. The
matching yields the $\mathbf{K}$-matrix as a function of scattering energy.  The $\mathbf{K}$-matrix
is a key quantity and other scattering observables can be obtained from it.

Finding and characterizing resonances is a major aspect of any
electron-molecule scattering study.  In this study our goal is to find
any higher lying resonances above the lowest and well-known $^2\Pi$
shape resonance.  In order to do this we use two quantities, the
eigenphase sum and time-delay, to find and fit the resonances.

The eigenphase sum fitting method is the standard method to detect
resonances in many studies. When the eigenphase sum is plotted against
the scattering energy resonances appear as sudden jumps by $\pi$ over
a small energy region \cite{haz79}.  Once located, the resonance
parameters can be found by fitting the eigenphase sum $\delta(E)$ to
the Breit-Wigner form
%\beq \delta(E) = \delta_{0}(E) +  \sum_{i=1}^m \tan^{-1} \frac{\Gamma_i}{2(E_i^r-E)} \eeq
\beq \delta(E) = \delta_{0}(E) + \tan^{-1} \frac{\Gamma}{2(E^r-E)}
\:,\label{eq:breitWigner} \eeq where $\delta_0(E)$ is the background
eigenphase, $E^r$ is the resonance position and $\Gamma$ is the width.
However, this method struggles to fit closely spaced and overlapping
resonances or ones near to a threshold.  The fitting of eigenphase sum
to the Breit-Wigner form is automatically done by the module RESON
\cite{jt31} in the UKRmol codes \cite{jt518}.

The above problems can be overcome in the time-delay method and it is,
therefore, the method of choice for detecting resonances in electron
collision with ionic targets, where there are large number of closely
spaced resonances.  The time-delay method was first proposed by Smith
\cite{smi60} and was first implemented in the UKRmol codes by Stibbe
and Tennyson \cite{jt188} through the module TIMEDEL \cite{jt227}.
Recently, it has been updated by Little {\it et al.} \cite{jt650}
and used to study electron collision with $N_2^+$ \cite{jt574}. If the
largest (first) eigenvalue of the time-delay matrix (i.e., the longest
time-delay) is plotted against scattering energy then the resonances
appear as Lorentzians. The TIMEDEL module automatically fits up to the
highest three eigenvalues as a function of energy ($q(E)$) to a
Lorentzian of the form: \beq q(E) =
\frac{\Gamma}{(E-E^r)^2+(\Gamma/2)^2} + bg(E) \:,\label{eq:lorenzian}
\eeq where $bg(E)$ is the background.

\section{Calculation \& Results} 

In this study we perform fixed-nuclei \rmat calculations for CO at the equilibrium
bond distance, $R_{eq}=2.1323$ a$_0$.  The molecular orbitals necessary for the
target and scattering calculations are obtained from MOLPRO \cite{molpro06}.
The scattering calculations are performed using the UK molecular \rmat\ codes
\cite{jt225,jt238}. These codes have been recently modernized and upgraded to
treat many different processes in electron and positron scattering with
molecules \cite{jt518}, and are called the UKRmol codes. Since neither 
the polyatomic implementation of UKRmol used here
nor MOLPRO can treat CO in its natural symmetry of $C_{{\infty}v}$ we use the
Abelian point group of $C_{2v}$.  Since identification of the target and
resonant scattering states in the $C_{2v}$ symmetry group to the
$C_{{\infty}v}$ is clear, we report these states using their natural symmetry
group. For the molecular orbitals (MOs) we use the $C_{2v}$ point group
designations.

We have performed extensive tests with respect to different target
and scattering models.  Our strategy had been to do these tests with
the smaller cc-pVDZ basis set in order to find a good and yet
computationally manageable model for the scattering calculations.
Then, we performed the final calculations using the bigger cc-pVTZ basis set
with the chosen model.

\subsection{Target calculations}

As described above, different scattering models involve use of
different types of target wave functions in the expansion in
Eq.~(\ref{eq:rmat}). In the SE and SEP model only the target ground
state represented at the SCF level is used. The SCF ground state
energy and dipole moment of CO is found to be $-112.74928$ E$_h$ and
$-0.23$ Debye, respectively, for the cc-pVDZ basis set. The HF
electronic ground state configuration of CO is given as $[(1a_1 -
5a_1)^{10}, (1b_1)^2, (1b_2)^2]$ in $C_{2v}$ symmetry or as $[(1\sigma
- 5\sigma)^{10}, (1\pi)^4]$ in $C_{{\infty}v}$ symmetry.

For the use in scattering calculations with the CC model, we performed
systematic CASSCF studies on the target using various active spaces for
the cc-pVDZ basis set. These active spaces are defined in Table
\ref{tab:casConfig}. The smallest and commonly used active space is
the full valence CAS (FVCAS), where all 10 valence electrons are
distributed among all 8 valence MOs, keeping the 4 core electrons
frozen.  We call this as CAS(10,8) and the electron configuration is
given as: $(1a_1 - 2a_1)^4 \, (3a_1 - 6a_1, 1b_1 - 2b_1, 1b_2 -
2b_2)^{10}$. The largest CASSCF calculation we performed is the
CAS(10,15) which, in addition to the valence MOs, also included the
lower $\sigma$ and $\pi$ molecular orbitals, formed from the 3s and 3p
atomic orbitals of C and O, in the active space. This calculation took
more than 30 hours for a sequential MOLPRO run on a 64-bit machine.
In order to do our final scattering calculations we choose, however,
the computationally more modest CAS(10,10) model.  This is because the
scattering calculation with any larger active space would become
unmanageably large with the larger cc-pVTZ basis set despite the use
of a specially-designed algorithm for Hamiltonian generation
\cite{jt180}.

A selected set of results from these CASSCF calculations for the
cc-pVDZ basis set is given in Table~\ref{tab:TargetEnergy}. These
calculations are performed for 40 target states, which includes the lowest
5 states from each space-spin symmetry.  Therefore, the MOs used in
the CC scattering calculations are the state-averaged CASSCF
(SA-CASSCF) orbitals having equal weights from each state. The table
compares the ground state energy (in $E_h$), vertical excitation
energies (in eV) to the lowest 9 states and ground state dipole moments
(in Debye) among different CASSCF models. As can be seen the relative
vertical excitation energies are fairly close to each other in these
CASSCF models. The table also includes the results from CAS(10,10)
calculation using the cc-pVTZ basis set. For the cc-pVTZ basis set we
made a 50 states SA-CASSCF calculation, where in addition to the
above said 40 target states we included 5 more states from each of the $^1A_1$
and $^3A_1$ symmetries. This was done in order to include a greater
number of $\Sigma^+$ target states in the CC scattering calculation, as
the $A_1$ state in $C_{2v}$ symmetry contains both $\Sigma^+$ and
$\Delta$ states. Having done this, we can see in the
Table~\ref{tab:TargetEnergy} that the second $^3\Sigma^+$ target state
becomes the ninth lowest excited state.  Available experimental
(adiabatic) excitation energies are also included in the table for
comparison. These values are derived by Nielsen \etal \cite{nie80}
from the spectroscopic constants of Huber and Herzberg \cite{hub79}.
Their spectroscopic assignments are given in the parenthesis. As can
be seen the experimental values are quite consistent with our vertical
excitation energies from the cc-pVTZ basis set.

\subsection{Asymptotic states}

As discussed below there are significant number of states of the CO$^-$
system which lie below the dissociation limit of CO into C($^3$P) and
O($^3$P).  Note that here and elsewhere we neglect spin-orbit effects
which lead to the splitting of these and other atomic terms values.

To help understand the high energy resonance structure in CO$^-$ we
performed an analysis of these states. Table~\ref{tab:asym} shows the
number and symmetry of the states which dissociate to the four bound
asymptotes, C($^3$P) + O$^-$($^2$P), C$^-$($^4$S) + O($^3$P), C($^1$D)
+ O$^-$($^2$P) and C$^-$($^2$D) + O($^3$P).  Based on the electron
affinities for O \cite{Blondel95} and C, \cite{Feld77} these
dissociation products are bound by 1.461, 1.262, 0.197 and 0.033 eV, 
respectively.

As detailed in Table~\ref{tab:asym}, there are 45 separate molecular
curves which lead asymptotically to states of CO$^-$ and
asymptotically lie below the C + O ground state.  In principle any of
these curves could be involved in dissociative attachment and have an
associated resonance signature at short internuclear separations, $R$.
To understand this situation better it was decided to perform a series
of electronic structure calculations to characterise these states. In
doing this we concentrate on the large $R$ region where the electronic
states are bound with respect to the CO ground state so as to avoid
spurious effects which can arise from performing bound electronic
structure calculation in the continuum  \cite{jt241}.

Calculations were performed using MOLPRO, which
provides a range of quantum chemical methods normally used for
computing bound electronic states.  Such calculations can be used to
describe resonant anionic states in the asymptotic region when it
crosses and lies below that of the neutral ground state. In the
resonant region it will require scattering methods (like the R-matrix
theory), or possibly stabilization procedures, to correctly describe
the potential energy curves (PECs).  

\begin{figure}
 \includegraphics{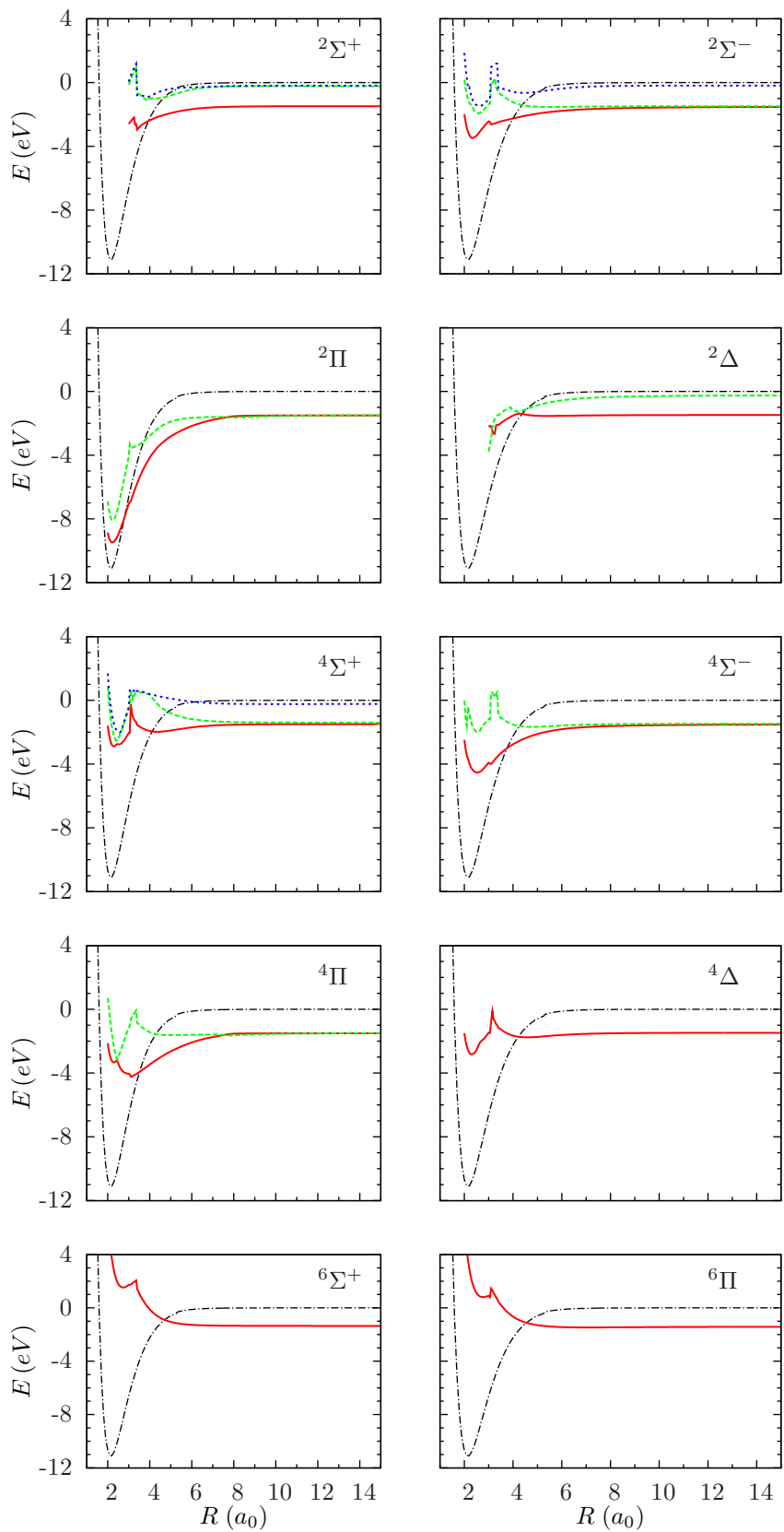}
\caption{\label{fig:pecs} CO$^-$ potential energy curves grouped by symmetry.
The dashed black curve is the X~$^1\Sigma^+$ CO ground state. The number of
curves are chosen using the numbers expected for each symmetry, see Table~\ref{tab:asym}.}
\end{figure}

In computing the PECs we use
the aug-cc-pVTZ basis set for C and O. The diffuse functions in the
augmented basis set are necessary to describe the anionic states. The
PECs are computed at the multi-reference configuration interaction
(MRCI) level of theory with Davidson correction. This method is
generally refered to as the MRCI+Q method. The Davidson correction is
an extrapolation method to the full-CI limit. The necessary molecular
orbitals are calculated from a state-averaged CASSCF(11,10)
calculation at the specified bond length. The active space for
the chosen CASSCF calculation is defined as: $(1a_1 - 2a_1)^4 \, (3a_1
- 6a_1, 1b_1 - 3b_1, 1b_2 - 3b_2)^{11}$.
The role of choice of CAS in the calculation of excited electronic
states has recently come under scrutiny \cite{jt623,jt632} but
was not explored here.  Figure~\ref{fig:pecs}
summarises our results.
The energies reported are from  MRCI+Q calculations and
 plotted with respect to the dissociation energy of the
neutral ground state CO molecule and converted to eV.

Figure~\ref{fig:pecs} suggests that the sextet states ($^6\Sigma^+$
and $^6\Pi$) as well as the $^4\Delta$ states are all repulsive at
short $R$.  For the other symmetries the calculations at least suggest
that there are states which could result in resonance features in the
10 - 15 eV region.  The only other exception to this is the $^2\Pi$
symmetry which, of course, shows the clear signature of the well-known
$^2\Pi$ shape resonance but the next curve appears to become a
quasi-continuum state at short bond lengths as its shape simply
mirrors that of the CO ground state curve.  Our calculations show that
the lowest $^2\Pi$ shape resonance correlates asymptotically with
C($^3$P) + O$^-$($^2$P) and that the lowest $^2\Sigma^+$ curves goes
asymptotically to C$^-$($^4$S) + O($^3$P).

\subsection{Scattering calculations}

All the reported scattering calculations use an \rmat sphere of radius $a = 10
\ a_0$.  The continuum orbitals, which represent the scattering electron, are
expanded in a basis of Gaussian-type functions centred on the centre-of-mass
of the target \cite{jt286}.  The continuum orbitals with partial waves $\ell
\leq 4$ are included in these calculation. These orbitals are Schmidt
orthogonalized to the target MOs and then all MOs are symmetrically
orthogonalized to each other. As noted above, in the SE model the target MOs
are SCF orbitals while in CC model these are SA-CASSCF orbitals. Only those MOs
that have eigenvalues, from the symmetrical orthogonalization, larger than a
deletion threshold of $10^{-7}$ are retained in the calculation. 

The results of the scattering calculations using various models and
the cc-pVDZ basis set are shown in the Table~\ref{tab:ResParameters}.
The SE model, unsurprisingly, finds only the $^2\Pi$ shape resonance
whose position and width are too large in comparison to the larger
models. All our CC calculations using the cc-pVDZ basis set also find
only one resonance, that is the lowest $^2\Pi$ shape resonance. Even
the 40 states CC calculation using a larger active space of CAS(10,11)
find only this resonance.We note that the inner region calculation for
A$_1$ symmetry with the larger CAS(10,11) took 5.5 days to finish in
comparison to using the CAS(10,10) which took only 5 hours.

In this study we do not report the results of SEP model.  Our previous
experience \cite{jt459} with SEP calculations showed that the
resonance parameters do not converge as the number of virtual MOs
included increases. This is because of the deteriorating  balance
between the target and scattering calculation. The scattering
calculation improves upon addition of more number of virtual MOs while
the target energy remains fixed in its ground state SCF
representation.  However, in an older SEP calculation on carbon
monoxide,  Morgan \cite{Morgan91} reported convergence of the lowest
$^2\Pi$ resonance position and width with respect to increase in
number of virtual MOs.  That calculation, however, was only for the
energy region below the first electronic excitation threshold. In SEP
there is also another problem due to the occurrence of the
non-physical pseudo-resonances \cite{jt474}. This problem arises
because of the fact that while this model includes polarization effect
it does not, however, include the excited target states in the
expansion in Eq.~(\ref{eq:rmat}).

Our \lq best' results are from the CC calculation which included 50 target states
represented by CAS(10,10) using the cc-pVTZ basis set. Since the \rmat method
is based on the variational principle, a lower value for resonance position
also means we have better approximation to the exact scattering wave function.
In this model the $^2\Pi$ resonance  position is found to be at 1.73 eV, which
is lower in comparison with all the  models tested using the cc-pVDZ basis set.  
We also find a new resonance in this model for the $^2A_1$ symmetry. Since it 
does not appear in the calculation for $^2A_2$ symmetry, we can assign it the
$^2\Sigma^+$ symmetry in $C_{\infty{v}}$ point group. 
 
\begin{figure} 
\includegraphics{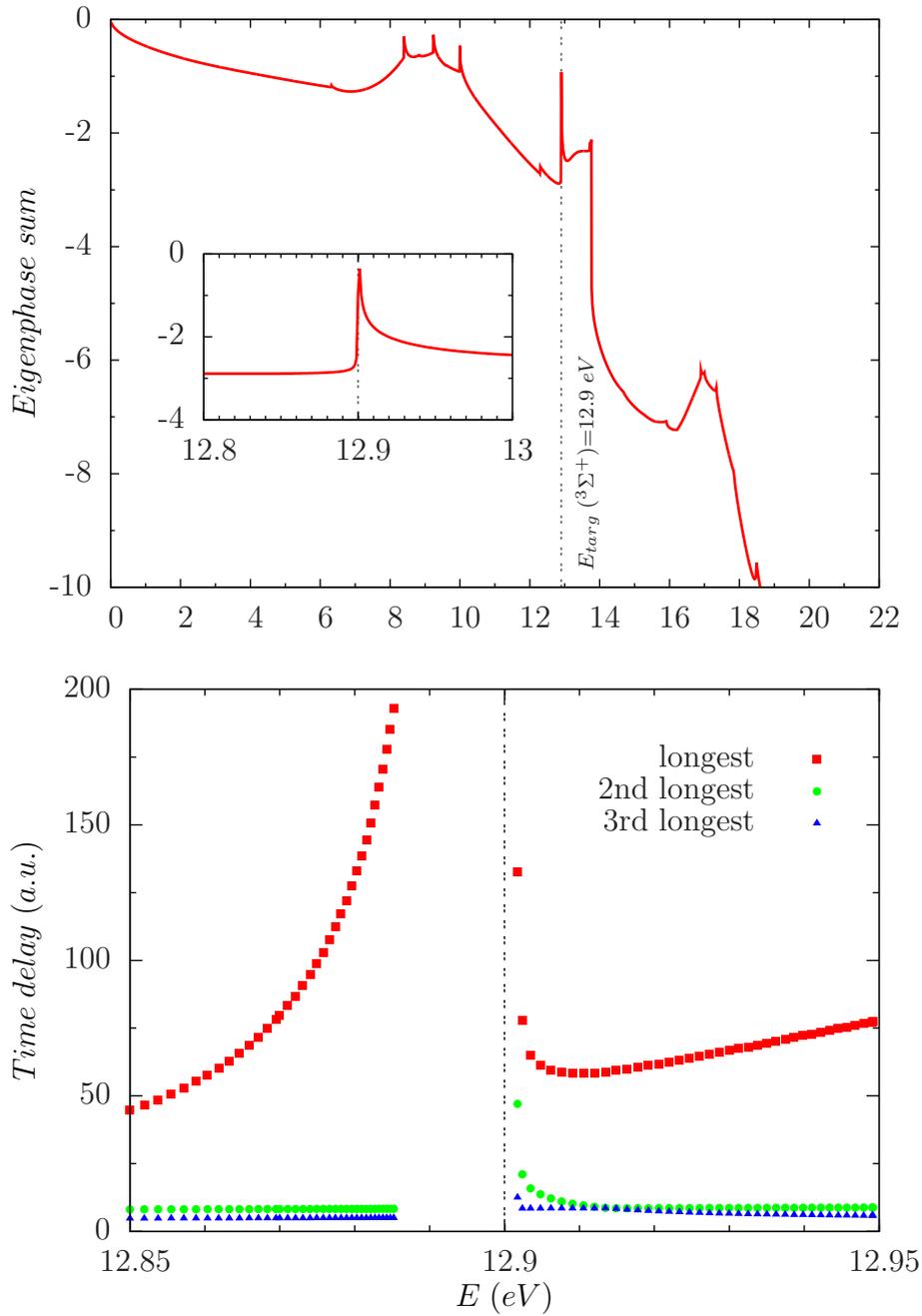}
\caption{ \label{fig:a1_reson} $^2A_1$ resonance: the upper
panel shows the eigenphase sum as a function of 
scattering energy from our best model (see text for details).
The inset shows the narrow Feshbach resonance at 12.9 eV, lying extremely 
close to the $2\; ^3\Sigma^+$ target state. The lower panel shows the 
time-delay plot in the resonance region.} 
\end{figure}

\begin{figure} 
\includegraphics{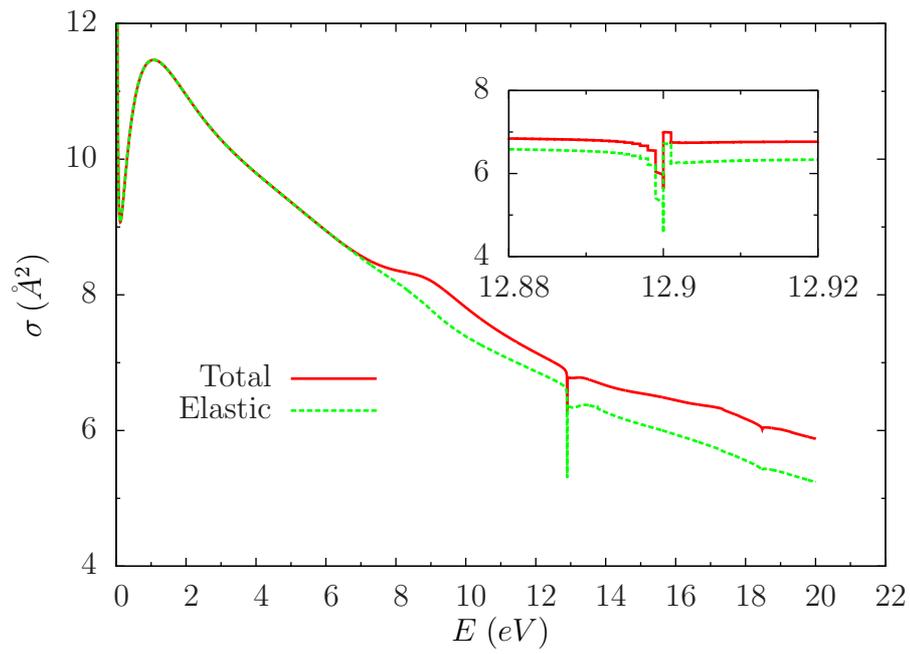}
\caption{\label{fig:a1_xsec} Total and elastic cross section for $^2A_1$ symmetry. The inset 
shows the effect of the Feshbach resonance to the cross sections.}
\end{figure}

The $^2\Sigma^+$ resonance is clearly seen in the eigenphase sum plot
for the $^2A_1$ symmetry in Fig.~\ref{fig:a1_reson}. The position of
this resonance is found to be 12.899988 eV with a width of 0.000525 eV
as fitted by RESON. The resonance position is extremely
close to the b~ $^3\Sigma^+$ threshold at 12.900787 eV.  In the
region of a threshold, the time-delay becomes infinite because the
scattering electron associated with the newly opening channel moves
with zero kinetic energy. This can be seen in the plot of time-delay
in Fig.~\ref{fig:a1_reson} where the time-delay diverges at 12.9 eV.
Therefore, we could not get the fitted resonance parameters from the
TIMEDEL module. Since this resonance has a very narrow width of 0.52 meV
and appears extremely close to the $2\,^3\Sigma^+$ target state at
12.9 eV, we therefore characterize it as a core-excited Feshbach
resonance with the $2\,^3\Sigma^+$ target state as its parent, to
which it is bound only by about 0.0008 eV.  The effect of the
resonance on the cross sections can be seen Fig.~\ref{fig:a1_xsec}.

\begin{figure} 
\includegraphics{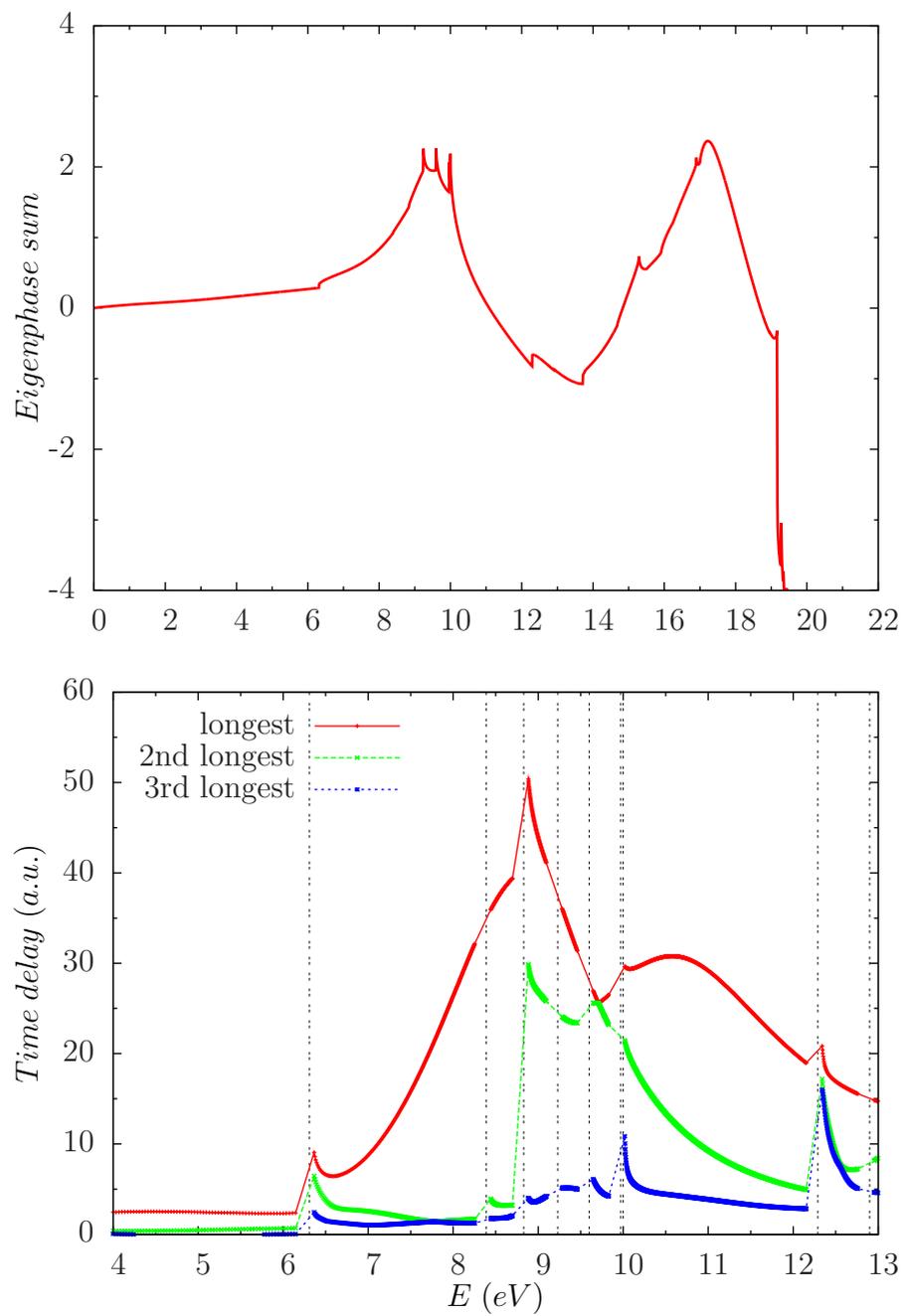}
\caption{\label{fig:a2_reson} Eigenphase sum plot (upper) and
 time-delay plot (lower) for $^2A_2$ symmetry. 
The broken vertical lines indicate the electronic excitation thresholds.
}
\end{figure}

\begin{figure} 
\includegraphics{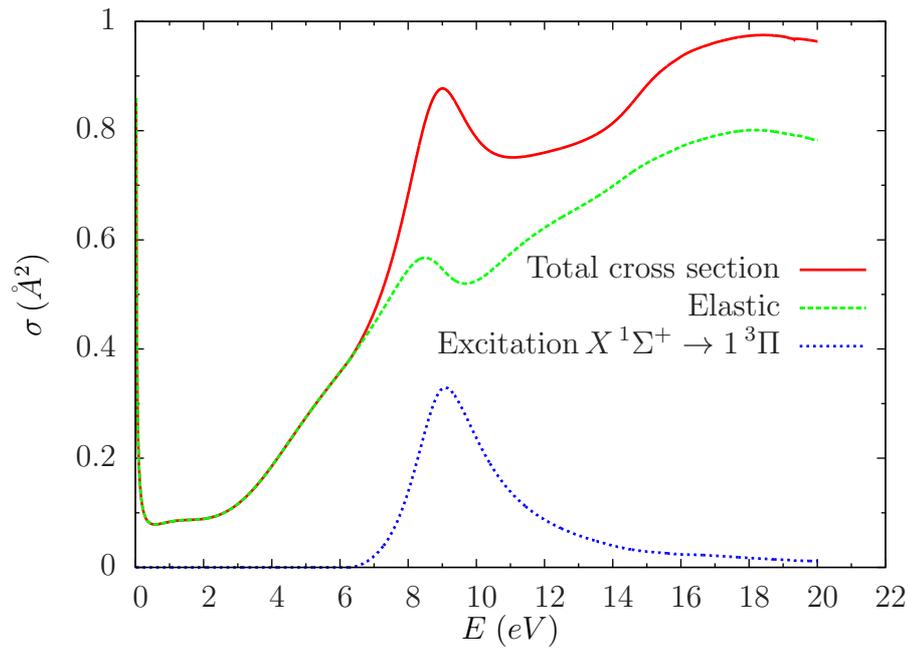}
\caption{\label{fig:a2_xsec} Total, elastic  and the dominant electron impact excitation cross
section of CO in the $^2A_2$ symmetry.} 
\end{figure}

We present the eigenphase sum and 
time-delay for $^2A_2$ symmetry from our best model  in  
Fig.~\ref{fig:a2_reson}. 
The plot for total, elastic and the dominant
electron impact excitation cross section of CO is given in
Fig.~\ref{fig:a2_xsec}. The cross section plot shows a broad peak at 9 eV.
However, the eigenphase sum and time-delay plots do not show any feature
associated with a resonance.  Neither the eigenphase sum and time-delay fitting
modules (RESON and TIMEDEL) find or fit any resonance for this symmetry.  We,
however, suspect that the peak structure at 9 eV will become a resonance at
larger bond distances. We have started doing \rmat calculations as a
function of bond distance with a view to investigating this and other aspects.
  
\begin{figure} 
\includegraphics{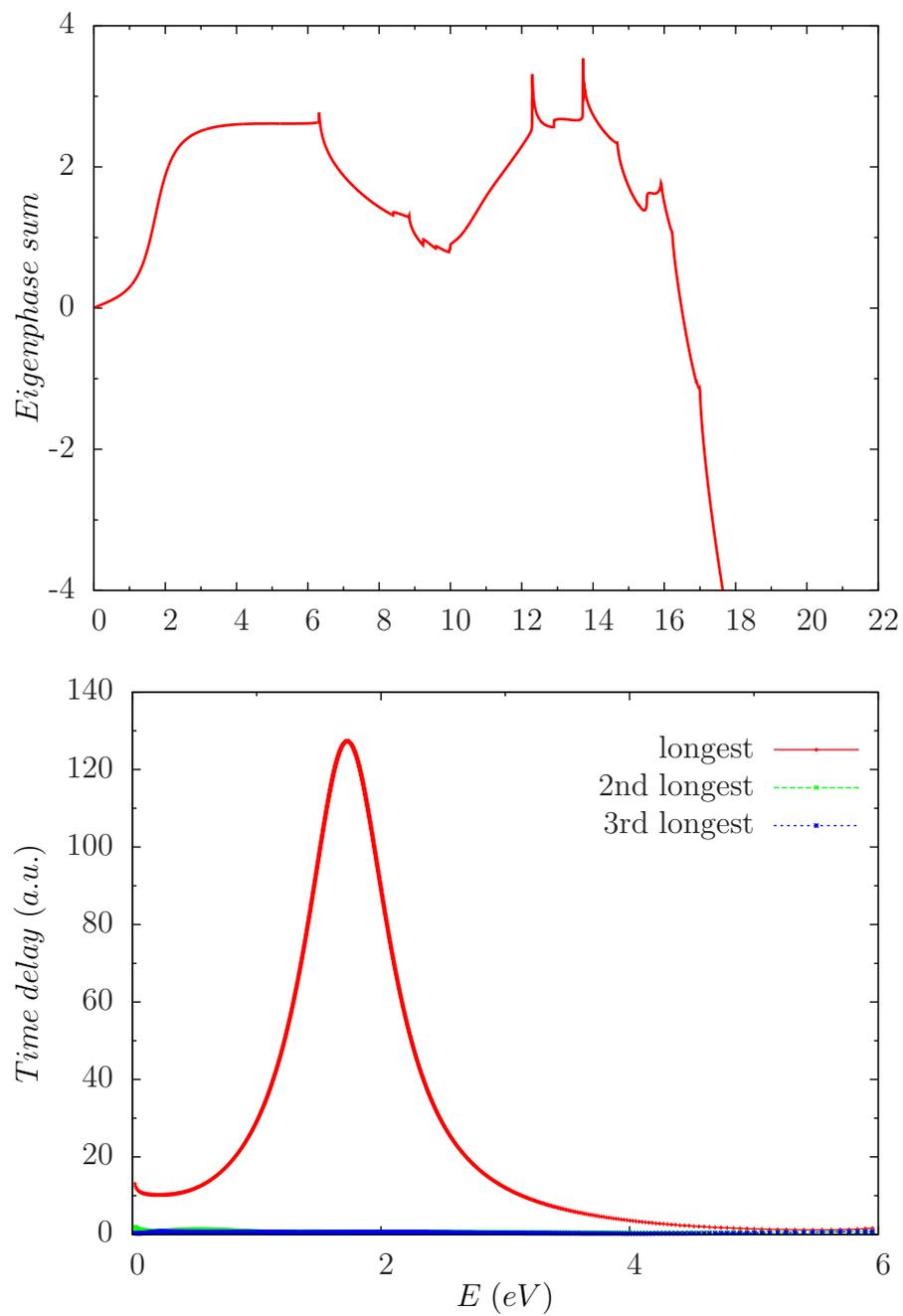}
\caption{ \label{fig:b1_reson} Eigenphase sum plot (upper) and
 time-delay plot (lower) for $^2B_1$ symmetry.}
\end{figure}

\begin{figure} 
\includegraphics{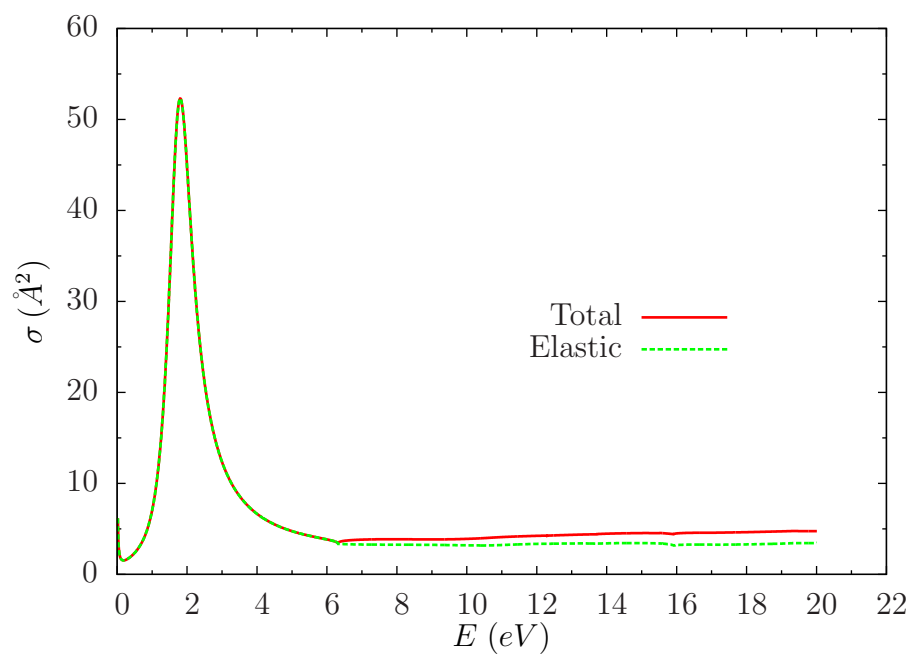}
\caption{\label{fig:pi_xsec} Total and elastic cross section for $^2B_1+^2B_2$ ($^2\Pi$) symmetry.} 
\end{figure}

The eigenphase sum and time-delay plots, from our best 
model, for the $^2B_1$ symmetry are presented in Fig.~\ref{fig:b1_reson};
these are identical to those obtained for the degenerate
$^2B_2$ symmetry calculation. The resonance
feature around 2 eV is fitted by both RESON and TIMEDEL to the same
values of position of 1.73 eV and width of 0.84 eV. The elastic and
total cross section due the $^2\Pi$ ($^2B_1+^2B_1$) symmetry is given
in Fig.~\ref{fig:pi_xsec}

\section{Discussion} 

The DEA experiments suggest that there are two $^2\Sigma^+$ resonances
between 10 eV and 11 eV \cite{rb65,ss70,cr71,ss71,75cth,hcs77,nn15}.
Our calculations detected only a single $^2\Sigma^+$ resonance at 12.9
eV. It is therefore worth discussing this difference.

There has long been experimental evidence that excited electronic
states of small molecules in the 10 to 15 eV region often support a
complicated set of Feshbach resonances \cite{Schulz73}. So far, theory
has only made a modest contribution to modelling and interpreting
these resonances. It is useful to consider the case of electron --
H$_2$ collisions where resonances in the 10 to 15 eV have been
well-studied. \rmat\ calculations on this system
\cite{jt90,jt101,jt215} mapped out resonances as a function of
internuclear separation to give resonance curves. However, these
calculations also found many ``features'' where the eigenphase sums
showed structures in form of resonance-like jumps, but that these
jumps were significantly smaller than one would expect from a
fully-formed resonance \cite{jt101}. Some of these features became
resonances as the internuclear separation was changed. Furthermore,
even for H$_2$, where with a two-electron target it is was possible to
perform full CI calculations, it was necessary to shift the resonance
positions to fully reproduce the observed behaviour \cite{jt208}. This
was done by identifying the parent associated with each Feshbach
resonance and then mapping this to highly accurate {\it ab initio}
curves which are, of course, available for H$_2$. Even here there is a
complication, as studies have shown, that Feshbach resonances
could often not be associated with a single parent state \cite{jt199}.
These H$_2^-$ resonance curves have recently been used for theoretical
studies of DEA and vibrational excitation of H$_2$ via these
high-lying resonances \cite{jt507,jt567}. We note that there was a
concerted attempt to map out such higher-lying Feshbach resonance in
water.  Here systematic studies of H$_2$O$^-$ resonances
\cite{hzm04,jt359,drw05,drw07} gave useful comparisons with experiment
but obtaining complete agreement with the observations remains more
difficult \cite{has11,13sha}. A similar methodology has been applied
to CO$_2$ \cite{sar11} and methane \cite{dsa15} in the 10 eV region.

For CO, the 10.04 eV $^2\Sigma^+$ resonance and its effect on the DEA 
was reported by Schulz \cite{Schulz73}. More recently, experimental DEA studies 
were undertaken by Nag and Nandy \cite{nn15} and Tian et. al. \cite{twx13}. These 
studies proved somewhat controversial regarding the nature of the resonances
involved. Whereas Nag and Nandy indicated the involvement of $^2\Sigma^+$
and $^2\Pi$ resonances in the DEA around 10-12~eV, Tian et al proposed that 
the DEA in the range 10-12~eV occurs through a coherent superposition of 
$^2\Pi$, $^2\Delta$ and $^2\Phi$ states at lower end of the energy range while 
at higher energies above 12.1 eV the resonant states in the superposition are 
changed to $^2\Sigma$, $^2\Delta$ and $^2\Phi$. 

Such higher lying resonances were reported in experimental studies of integral 
cross sections in vibrationally elastic transitions in CO \cite{jz96a,jz96}. 
These studies suggest that there were  several $^2\Sigma^+$ 
Feshbach resonances in the region above 10.04 eV  associated with the 
b~$^3\Sigma^+$ and higher lying parent states.

Returning to the present results we find a $^2\Sigma^+$ symmetry resonance
0.0008 eV below the second $^3\Sigma^+$ target state. This state,
which is known from experiment and labelled the b~$^3\Sigma^+$ state, has
been the subject of \rmat\ studies which used electron collisions
with CO$^+$ to characterize high-lying excited states of CO \cite{jt189,jt373}.
These studies suggest that the vertical excitation energy of the 
 b~$^3\Sigma^+$ state is in the region of 10.2 to 10.4 eV, in good
agreement with the observed result which places this at 10.4 eV \cite{ts72}.
It would therefore seem likely the the  $^2\Sigma^+$ Feshbach resonance
we detect lies somewhere in this region. Once nuclear motion effects
due to zero point energy, which is about 0.27 eV for CO, and other
effects are taken into account it would seem likely that this
resonance is responsible for the 10.2 eV DEA feature.
This would be in agreement with the stabilization calculations
of  Pearson and  Lefebvre-Brion \cite{pl76} who also only
identified a single resonance in the 10 eV region,
and also in line with Nag and Nandy's \cite{nn15} assertion
that a $^2\Sigma^+$ resonance is involved in DEA in this region.

A recent, detailed {\it ab initio} study by V{\'a}zquez {\it et al}
\cite{val09} of electronically excited states of CO demonstrate just
how complicated the curves are as function of internuclear separation
in this region.  Unfortunately V{\'a}zquez {\it et al} did not
consider states of $^3\Sigma^+$ symmetry so cannot be used to inform
our study.

So if we have successfully identified the lower of the two resonance
features, what about the resonance responsible for DEA at 10.9 eV?
There would appear to be two possibilities here. We see 
a number of features which could not be fully characterized as
resonances in our calculations. It is possible that as the bond length
increases one of these becomes a proper resonance which correlate
with one of the many CO$^-$ asymptotic states we identify and hence
can lead to DEA. However, it is more likely that this resonance is associated
with a target state which lies even higher than the b~$^3\Sigma^+$ state.
Table~\ref{tab:TargetEnergy} only considers the 9 lowest electronically
excited states of CO. In fact our calculation uses 50 states of CO
but the higher states give increasing unrealistic representations of
the physical target states; indeed many of them lie above the CO
ionization threshold. It would appear that to make progress it would be
necessary to design a model with an increased number of actual
target states (as opposed to pseudo-states) explicitly included in the model.

\section{Conclusion}

We have performed an initial \rmat\ study to try and identify the
resonance states responsible for dissociative electron attachment
(DEA) in the electron -- CO system. We identify a very narrow
$^2\Sigma^+$ Feshbach resonance which would appear to be the feature
which causes DEA at about 10.2 eV. In future work we will study this
resonance as function of internuclear separation, which should allow a
full model of the DEA process to be built. The narrowness of this
resonance will make the nuclear motion part of this model rather
straightforward since non-adiabatic effects can almost certainly be
neglected.

Our calculations failed to identify  further higher-lying resonances.
It is likely that such resonance are associated with parent target states
that is not well-represented in our model. The  higher-lying
electronic states in CO are increasingly Rydberg-like \cite{jt189,val09}
and therefore difficult to represent using standard target models.
Furthermore,
including such Rydberg states in a scattering calculation remains
very challenging \cite{jt585} and will probably require further work
on the methodology we use for such scattering calculations, for example
the routine use of extended \rmat\ spheres. Such work is currently
being undertaken as part of the development of the B-spline-based UKRMol+
codes \cite{Masin}; initial results on the much simpler electron -- BeH
collision system have demonstrated the methods ability to deal with
very diffuse target states and include a comprehensive treatment
of target electron correlation \cite{Darby}.

\section*{Acknowledgement}

AD gratefully acknowledges UCL for the use of its computational resources.
We thank Lorenzo Lodi for providing a procedure to generate the results
given in Table 3.

\vskip 0,2cm

All authors contributed equally to this work.
\bibliographystyle{epj}
%\bibliography{journals_phys,jtj,refs,rmat}

\clearpage
\newpage
\onecolumn

\begin{table}
 \caption{The active space configurations used in the target CASSCF calculations.
                  The molecular orbitals are labelled using $C_{2v}$ point group.}
  \label{tab:casConfig}
\begin{footnotesize}
\begin{tabular}{lc}
\hline\noalign{\smallskip}
%=====================================================================================================
\multicolumn{1}{c}{CASSCF models} & \multicolumn{1}{c}{Configurations}  \\
%=====================================================================================================
\hline\noalign{\smallskip}
 CAS(10,8) & $(1a_1 - 2a_1)^4 \, (3a_1 - 6a_1, 1b_1 - 2b_1, 1b_2 - 2b_2)^{10}$ \\
 CAS(10,9) & $(1a_1 - 2a_1)^4 \, (3a_1 - 7a_1, 1b_1 - 2b_1, 1b_2 - 2b_2)^{10}$ \\
CAS(10,10) & $(1a_1 - 2a_1)^4 \, (3a_1 - 6a_1, 1b_1 - 3b_1, 1b_2 - 3b_2)^{10}$ \\
CAS(10,11) & $(1a_1 - 2a_1)^4 \, (3a_1 - 7a_1, 1b_1 - 3b_1, 1b_2 - 3b_2)^{10}$ \\
CAS(10,12) & $(1a_1 - 2a_1)^4 \, (3a_1 - 8a_1, 1b_1 - 3b_1, 1b_2 - 3b_2)^{10}$ \\
CAS(10,13) & $(1a_1 - 2a_1)^4 \, (3a_1 - 9a_1, 1b_1 - 3b_1, 1b_2 - 3b_2)^{10}$ \\
CAS(10,14) & $(1a_1 - 2a_1)^4 \, (3a_1 - 8a_1, 1b_1 - 4b_1, 1b_2 - 4b_2)^{10}$ \\
CAS(10,15) & $(1a_1 - 2a_1)^4 \, (3a_1 - 9a_1, 1b_1 - 4b_1, 1b_2 - 4b_2)^{10}$ \\
%=====================================================================================================
\noalign{\smallskip}\hline
\end{tabular}
\end{footnotesize}
\end{table}

\clearpage
%\begin{sideways}
%\makebox[0.08]{
\begin{table}
\setlength\tabcolsep{1.5pt}
\caption{The ground state energy (in E$_h$), the lowest 9 vertical excitation energies (in eV) and ground state dipole moments ($\mu$ in D) of CO calculated     using varying active spaces with cc-pVDZ and using CAS(10,10) with cc-pVTZ basis sets.      See text for details. Experimental values derived by Nielsen \etal\ \cite{nie80} from the spectroscopic constants of Huber and Herzberg \cite{hub79} are given for comparison.}
  \label{tab:TargetEnergy}
\begin{tabular} {lcccccccccc} 
%\begin{footnotesize}
\hline\noalign{\smallskip}
%------------------------------------------------------------------------------------------
\multicolumn{1}{c}{State} &   \multicolumn{8}{c}{cc-pVDZ} & \multicolumn{1}{c}{cc-pVTZ} & \multicolumn{1}{c}{Expt} \\ \cline{2-9} 
   & CAS(10,8) & CAS(10,9) & CAS(10,10) & CAS(10,11) & CAS(10,12) & CAS(10,13) & CAS(10,14) & CAS(10,15) & CAS(10,10) \\
%-------------------------------------------------------------------------------------------
\noalign{\smallskip}\hline
% State                                     cc-pVDZ                                                                          cc-pVTZ     Expt.
$X\,^{1}\Sigma^+$ & -112.85537 & -112.86388 & -112.89473 & -112.92428 & -112.92723 & -112.93840 & -112.95161 & -112.95146 & -112.85655                        \\
$1\,^{3}\Pi$      & 6.49       & 6.52       & 6.40       & 6.39       & 6.38       & 6.56       & 6.38       & 6.41       & 6.31       & 6.32 ($a\,^{3}\Pi$)  \\
$1\,^{3}\Sigma^+$ & 8.69       & 8.69       & 8.79       & 8.83       & 8.80       & 8.78       & 8.80       & 8.72       & 8.39       & 8.51 ($a'\,^{3}\Sigma^+$)  \\
$1\,^{1}\Pi$      & 9.12       & 9.12       & 9.19       & 9.22       & 9.16       & 9.33       & 9.15       & 9.02       & 8.83       & 8.51 ($A\,^{1}\Pi$)  \\
$1\,^{3}\Delta$   & 9.62       & 9.64       & 9.76       & 9.81       & 9.80       & 9.80       & 9.80       & 9.74       & 9.23       & 9.36 ($d\,^{3}\Delta$)  \\
$1\,^{3}\Sigma^-$ & 10.00      & 10.02      & 10.15      & 10.20      & 10.18      & 10.19      & 10.16      & 10.12      & 9.60       & 9.88 ($e\,^{3}\Sigma^-$)  \\
$1\,^{1}\Sigma^-$ & 10.37      & 10.42      & 10.54      & 10.59      & 10.59      & 10.60      & 10.58      & 10.54      & 9.97       & 9.88 ($I\,^{1}\Sigma^-$)  \\
$1\,^{1}\Delta$   & 10.41      & 10.46      & 10.57      & 10.62      & 10.61      & 10.64      & 10.59      & 10.57      & 10.00      & 10.23 ($D\,^{1}\Delta$)  \\
$2\,^{3}\Pi$      & 12.84      & 12.91      & 12.65      & 12.89      & 12.89      & 12.88      & 12.78      & 12.75      & 12.29 \\   
$2\,^{1}\Pi$      & 14.36      & 14.40      & 14.24      & 14.45      & 14.41      & 14.48      & 14.30      & 14.24      & \\   
$2\,^{3}\Sigma^+$ &            &            &            &            &            &              &          &            &12.90\\

$\mu$ & 0.514 & 0.452 & 0.234 & 0.071 & 0.045 & 0.158 &  0.043 &  0.240 & 0.291 &  0.122 \\
%------------------------------------------------------------------------------------------
\noalign{\smallskip}\hline
%\end{footnotesize}
\end{tabular}
\end{table}

\begin{table}
  \caption{Molecular curves correlating with the bound asymptotic states of CO$^-$;
given are both the number of molecular states correlating with each dissociation
product, $N$, and their symmetries. 2 $^2\Sigma^-$ means two states of $^2\Sigma^-$
symmetry and so forth.
Binding energies, $E_b$, are given relative to C($^3$P) + O($^3$P).}
  \label{tab:asym}
\begin{tabular}{lcrl} 
\noalign{\smallskip}\hline
\multicolumn{1}{c}{Product} & \multicolumn{1}{c}{$E_b$ / eV} & $N$&\multicolumn{1}{l}{Symmetries} \\
%====================================================================================================
\noalign{\smallskip}\hline
C($^3$P) + O$^-$($^2$P)& 1.461 &12& 
$^2\Sigma^+$, 2 $^2\Sigma^-$, 
2 $^2\Pi$, 
$^2\Delta$, 
$^4\Sigma^+$, 2 $^4\Sigma^-$, 
2 $^4\Pi$,  
$^4\Delta$\\
C$^-$($^4$S) + O($^3$P)& 1.262 & 6& 
$^2\Sigma^+$,
$^2\Pi$, 
$^4\Sigma^+$,
$^4\Pi$,
$^6\Sigma^+$,
$^6\Pi$\\
C($^1$D) + O$^-$($^2$P)& 0.197 &9&
2 $^2\Sigma^+$, $^2\Sigma^-$,
3 $^2\Pi$,
2 $^2\Delta$,
$^2\Phi$\\
C$^-$($^2$D) + O($^3$P)& 0.033 &18&
2 $^2\Sigma^+$,$^2\Sigma^+$,
3 $^2\Pi$, 
2 $^2\Delta$, 
$^2\Phi$,
2 $^4\Sigma^+$, $^4\Sigma^-$,
3 $^4\Pi$,
2 $^4\Delta$, 
$^4\Phi$\\
\noalign{\smallskip}\hline
\end{tabular}
\end{table}

 \clearpage                                                                     
\newpage
%%%%%%%%%%%%%%%%%%%%%%%%%%%%%%%%%%%%%%%%%%%%%%%%%%%%%%%%%%%%%%%%%%%%%%%%%%%%%%%%%%%%%%%%%%%%%%%%%%%%
%\begin{scriptsize}
\begin{table}
  \caption{Positions (and widths) of the detected $^2\Pi$ and $^2\Sigma^+$ resonances 
              computed using different scattering models. The label CAS(10,8)/cc40 represents a close-coupling calculation
              with 40 target states represented by CAS(10,8) model. None of our calculations using cc-pVDZ basis set detected
              the $^2\Sigma^+$ resonance. All quantities are in eV.}
  \label{tab:ResParameters}
\begin{tabular}{cccc} 
\noalign{\smallskip}\hline
%=====================================================================================================
\multicolumn{1}{c}{Basis sets} & \multicolumn{1}{c}{Models} & \multicolumn{1}{c}{$^2\Pi$ resonance} & \multicolumn{1}{c}{$^2\Sigma^+$ resonance} \\
%====================================================================================================
\noalign{\smallskip}\hline
cc-pVDZ  & SCF/SE           & 3.50~(1.96)\\
         & CAS(10,8)/cc40   & 2.01~(0.83)\\
         & CAS(10,10)/cc40  & 2.12~(0.91)\\
         & CAS(10,10)/cc50  & 1.95~(0.81)\\
         & CAS(10,11)/cc40  & 2.20~(0.95)\\
%-----------------------------------------------------------------------------------------------------
cc-pVTZ  & CAS(10,10)/cc50  & 1.73~(0.84) &  12.899988~(0.000525)  \\
%=====================================================================================================
\noalign{\smallskip}\hline
\end{tabular}
\end{table}

%\end{scriptsize}
%%%%%%%%%%%%%%%%%%%%%%%%%%%%%%%%%%%%%%%%%%%%%%%%%%%%%%%%%%%%%%%%%%%%%%%%%%%%%%%%%%%%%%%%%%%%%%%%%%%%%

\clearpage

\newpage

\clearpage
\clearpage
\clearpage
\clearpage

\end{document}